\documentclass[a4paper, 11pt]{article}

\usepackage{graphicx}
\usepackage{epstopdf,epsfig}
\usepackage{newtxtext}
\usepackage{newtxmath}
\usepackage{natbib}
\usepackage{hyperref}
\usepackage{subcaption}
\usepackage{orcidlink}
\graphicspath{{figs/}}
\hypersetup{
	colorlinks = true,
	urlcolor   = blue,
	citecolor  = black,
}

\newcommand{\RomanNumeralCaps}[1]

\newcommand{\rev}[1]{#1}

\newcommand{\av}[1]{\langle {#1} \rangle}

\newcommand{\avl}[1]{ \widetilde{#1}}

\title{On small-scale and large-scale intermittency of Lagrangian statistics in canopy flow}

\usepackage[margin=2.7cm]{geometry}
\usepackage{orcidlink}
\author{Ron Shnapp\footnote{ronshnapp@gmail.com} \, \orcidlink{0000-0001-7495-8420} \\[.5em] \normalsize{Department of Physics of Complex Systems, Weizmann Institute of Science, Israel}}
\date{}

\begin{document}
\maketitle

\begin{abstract}
	
	\noindent The interaction of fluids with surface-mounted obstacles in canopy flows leads to strong turbulence that dominates dispersion and mixing in the neutrally stable atmospheric surface layer. This work focuses on intermittency in the Lagrangian velocity statistics in a canopy flow, which is observed in two distinct forms. The first, small scale intermittency, is expressed by non-Gaussian and not self-similar statistics of the velocity increments. The analysis shows an agreement in comparison with previous results from homogeneous isotropic turbulence (HIT) using the multifractal model, extended self-similarity, and velocity increments' autocorrelations. These observations suggest that the picture of small-scale Lagrangian intermittency in canopy flows is similar to that in HIT, and therefore, they extend the idea of universal Lagrangian intermittency to certain inhomogeneous and anisotropic flows. Second, it is observed that the RMS of energy increments along Lagrangian trajectories depend on the direction of the trajectories' time-averaged turbulent velocity. Subsequent analysis suggests that the flow is attenuated by the canopy drag while leaving the structure function's scaling unchanged. This observation implies the existence of large-scale intermittency in Lagrangian statistics. Thus, this work presents a first empirical evidence of intermittent Lagrangian velocity statistics in a canopy flow that exists in two distinct senses and occurs due to different mechanisms.
\end{abstract}



\section{Introduction}\label{sec:intro}

Turbulent flows are often characterized by bursts of activity amongst long quiescent periods, and thus, they are said to be intermittent. Intermittency can occur in turbulence in two different forms. The first is called small-scale intermittency, and it was first reported by \cite{Batchelor1949}, and reviewed by \cite{Frisch1995, Tsinober2009}. Small-scale intermittency is evident in statistics of velocity differences, both in the Eulerian and the Lagrangian frames, since their probability distribution functions (PDFs) develop increasingly heavier tails as the scale of separation is reduced (e.g. \cite{Kailasnath1992, Arneodo2008}). Despite numerous models that have been suggested, a comprehensive theory for small-scale intermittency is still missing (e.g. \cite{She1994, Elsinga2020}), and yet, it is believed to be a universal feature of high Reynolds number turbulence. 
The second kind of intermittency is termed large-scale intermittency, and it may occur due to variability of the flow at low frequencies. For example, transitions between the turbulent and non-turbulent states occur in jets or in transitional pipe flows  \citep{Corrsin1943, Wygnanski1973}, mesoscales can change local turbulence parameter in the atmospheric boundary-layer \citep{Muchinski2004}, and strong large-scale velocity and temperature fluctuations can occur in stratified flows \citep{Feraco2018}. This work focuses on flows that are typical of the atmospheric surface layer, so-called canopy flows. In these flows, a fluid flow interacts with large surface-mounted obstacles, leading to high turbulence intensities. Furthermore, turbulence in canopies is said to be non-local since a significant fraction of turbulent kinetic energy is produced at the top of the obstacles and is then transported into the canopy layer itself \citep{Finnigan2000}. The non-local character of the turbulence in canopy flows leads to large-scale intermittency inside the canopy, which is expressed by a velocity skewness, sparse extreme events of momentum and scalar fluxes or time-varying H\"{o}lder exponents (e.g. \cite{Finnigan1979, Gao1989, Louka2000, Keylock2020}). Therefore, canopy flows provide a fruitful ground for observing the two phenomena in conjunction, which is the aim of this work.

Intermittency in turbulence was studied previously mostly in the Eulerian framework, yet the advent of technological advances of the 2000s enabled empirical investigations in the Lagrangian framework as well (as reviewed by \cite{Toschi2009}). Previous Lagrangian studies have revealed the existence of anomalous scaling of velocity differences \citep{Chevillard2003, Arneodo2008,Benzi2010,Huang2013}, have examined local flow features associated with extreme events \citep{Liberzon2012, Xu2014, Watteaux2019} and proposed modeling strategies \citep{Wilczek2013, Bentkamp2019}. These works focused on homogeneous isotropic turbulent flows and, inevitably so, focused on small-scale intermittency. \rev{Indeed, there is an absence of Lagrangian studies focusing on intermittency in inhomogeneous flows. In addition, \cite{Blum2010} showed that the Lagrangian structure functions depended on the magnitude of the instantaneous large-scale velocity in an oscillating grids experiment. Other than that, there is a lack of studies that focus on large-scale intermittency in the Lagrangian framework.} In particular, there are no empirical investigations of intermittency in Lagrangian statistics in canopy flows despite its importance to Lagrangian stochastic models with applications for dispersion and mixing in the environment \citep{Wilson1996, Reynolds1998, Duman2016, Shnapp2020c, Viggiano2020, Keylock2020}.

This work presents an analysis of Lagrangian statistics in a canopy flow using empirical results from a recent wind-tunnel experiment \citep{Shnapp2019a}. The existence of small-scale intermittency is demonstrated in Sec.~\ref{sec:multifractal}, and the results are compared to previous studies from HIT flows. Both qualitative and quantitative agreement is observed, which supports the idea of the universality of small-scale Lagrangian intermittency suggested by \cite{Arneodo2008}. Then, in Sec.~\ref{sec:activity-momentum}, it is demonstrated through conditional statistics that large-scale intermittency existed in the canopy flow as well, and that although it affected the energetics of trajectories, it did not affect the scaling laws of structure functions.

\section{Methods}\label{sec:methods}

Lagrangian trajectories in a canopy flow were analyzed using the results of a wind-tunnel, 3D particle tracking velocimetry (3D-PTV) experiment. The full experimental details are given in \cite{Shnapp2019a}, and Lagrangian statistics were analyzed in \cite{Shnapp2020c}. For brevity, only the information relevant to this work shall be repeated here.

\begin{figure}
	\centering
	\includegraphics[height=4.0cm]{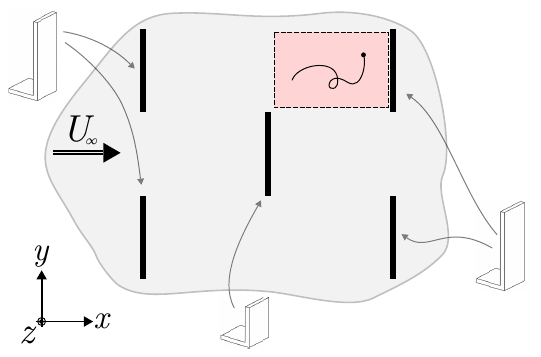}
	\caption{A schematic diagram of several canopy obstacles in top view in the wind tunnel. The measurement volume is highlighted in red upstream of a tall canopy obstacle. \label{fig:setup}}
\end{figure}

The experiment was conducted in the environmental wind-tunnel laboratory at the Israel Institute for Biological Research (IIBR), which features a 14 meters long open wind-tunnel with a $2 \times 2$ $\mathrm{m}^2$ cross-sectional area. We used a double-height staggered canopy layout, in which flat plates of height $H$ and $\frac{1}{2}H$ were placed in consecutive rows ($H=100\mathrm{mm}$). The plates were thin, their width was $\frac{1}{2}H$, and the spacing between the rows was $\frac{3}{4}H$, as shown in the sketch in Fig.~\ref{fig:setup}. The canopy frontal area density was $\lambda_f = A_f / A_T=\frac{9}{16}$, (where $A_f$ is the element's frontal area and $A_T$ is the lot area of the canopy layer), which categorizes our canopy as moderately dense. The wind velocity was $U_\infty=2.5 \, \mathrm{m\,s^{-1}}$, corresponding to a Reynolds number of $\mathrm{Re}_\infty = U_\infty H / \nu = 1.6 \times 10^4$ ($\nu$ is the kinematic viscosity). We recorded the trajectories using a real-time image analysis extension of the 3D-PTV method described in \cite{Shnapp2019a}. The PTV algorithms and the analysis were applied using the \cite{openptv} open-source software and the Flowtracks package by \cite{Meller2016}. In this work, $x$ is the streamwise direction, $y$ is the horizontal spanwise direction, and $z$ is perpendicular to the bottom wall directed upwards, where $z=0$ corresponds to the bottom wall.

This work is focused on a subset of trajectories that were recorded in a small sub-volume of space. The sub-volume had a length of $\frac{3}{4} H$, width of $\frac{1}{2}H$, and it was situated at the top of the canopy layer, $0.9 <\frac{z}{H}<1.1$ (this is sub-volume b3 in \cite{Shnapp2020c}, and it is shown in Fig.~\ref{fig:setup}). The RMS of velocity fluctuations was $u'=0.47 \, \mathrm{m\,s^{-1}}$, the mean dissipation rate was estimated as $\epsilon = 0.25 \, \mathrm{m^2 \,s^{-3}}$, the Kolmogorov length scale was $\eta = 0.34 \, \mathrm{mm} \approx \frac{1}{290}H$, and the Taylor microscale Reynolds number was $\mathrm{Re}_\lambda = 440$. Furthermore, the Lagrangian streamwise velocity decorrelation timescale was $T=54 \, \mathrm{ms}$, estimated by fitting Lagrangian velocity autocorrelation function. Notably, the decorrelation timescale varied for each velocity component, and the Lagrangian integral timescale $T_L$ is not trivial to define, so $T$ shall be used as a proxy for $T_L$ for simplicity (see \cite{Shnapp2020c} for a detailed discussion).


\section{Results}

\subsection{Lagrangian velocity increments and small-scale intermittency}\label{sec:multifractal}

In the following section, the focus is put on small-scale intermittency. The Lagrangian temporal velocity increment, defined as
\begin{equation}
\Delta_\tau v_i (t_0) \equiv v_i(t_0 + \tau) - v_i(t_0)
\label{eq:velocity_diff}
\end{equation}
where $\tau$ is the time lag, is widely used to study velocity statistics at different scales. Here, we use statistics of $\Delta_\tau v_i$ to show the existence of, and to analyze, small-scale intermittency in the canopy flow. Note that assuming stationarity of the flow in the wind tunnel, statistics are reported for different trajectories with different $t_0$, namely, we average over $t_0$. \rev{Furthermore, statistics of $\Delta_\tau v_i$ are assumed stationary in the range of $\tau$ considered here, due to the local homogeneity that we have shown in~\cite{Shnapp2020c}.}

\begin{figure}
	\centering
	\includegraphics[height=5cm]{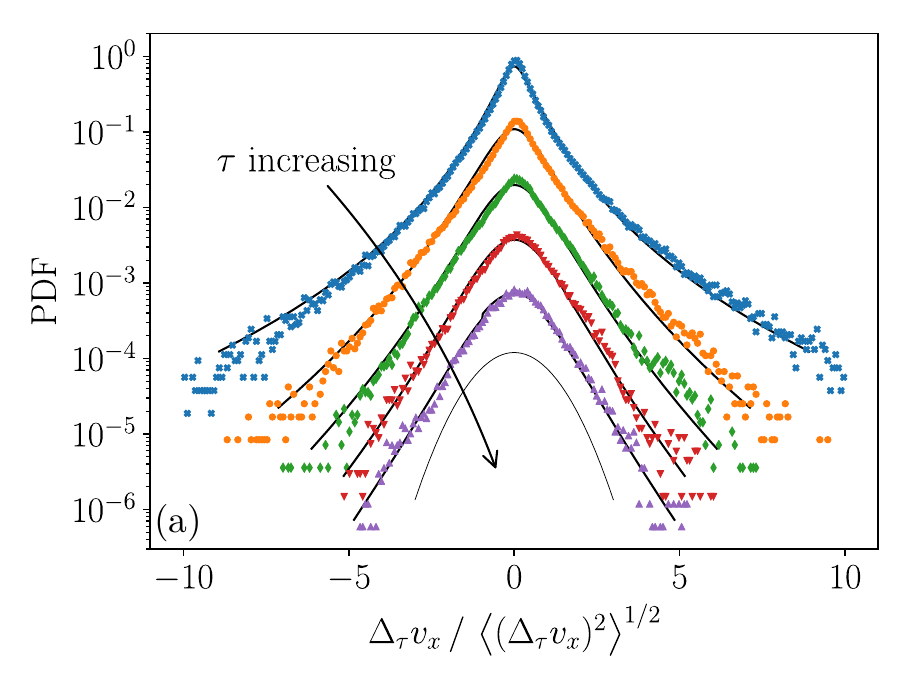}
	\includegraphics[height=5cm]{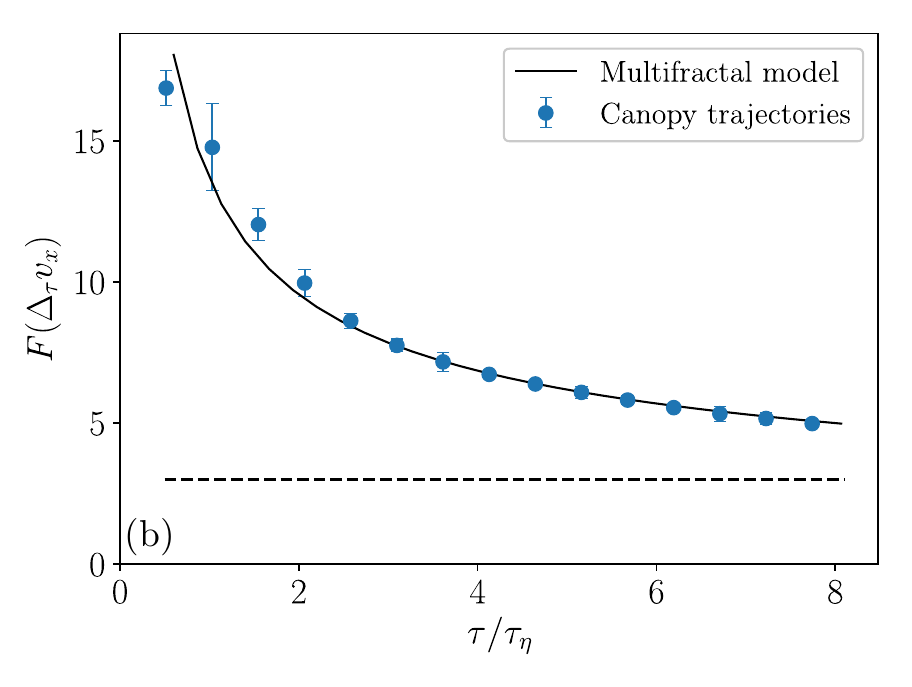}
	\caption{(a) - standardized PDFs of Lagrangian temporal velocity increments at $\tau/\tau_\eta \in \{0.3, 3, 5, 8, 11\}$, translated vertically; symbols correspond to the empirical canopy flow data, black lines stand for the multifractal model, and a Gaussian PDF is shown as a thin gray line at the bottom. (b) - The flatness of Lagrangian velocity increments plotted against the time lag; the results of the multifractal model shown as a black line, and the Gaussian value $F=3$ is marked by a dashed line.}
	\label{fig:velocity_increments}
\end{figure}

The PDFs, $P(\Delta_\tau v_x)$, for trajectories from the canopy flow experiment are shown in Fig.~\ref{fig:velocity_increments} (a) as symbols for five values of $\tau$. The PDFs were translated vertically for better visualization. 
\rev{The figure shows that despite the average flow velocity and its inhomogeneity, the velocity increments are zero averaged.} The figure also shows that as the time lag is reduced the tails of the PDFs become wider, showing that at smaller scales there is a higher probability for observing extreme events. In addition, the flatness coefficient of the velocity differences is plotted in Fig~\ref{fig:velocity_increments} (b) against $\tau/\tau_\eta$. The empirical data is shown as symbols, error bars represent the range obtained using bootstrapping with 5 sub-samples of the data, and the Gaussian value of $F=3$ is shown as a dashed line. \rev{Due to the available volume of the data the flatness at small $\tau$ is underestimated in our analysis; and still, at small $\tau$ the flatness is high, reaching roughly 17, and as the time lag grows it reduces monotonously and reaches down to $F\approx5$.} In the Kolmogorov similarity theory, dimensional analysis predicts that moments of the velocity difference scale with the time lag as $\av{(\Delta_\tau v_i)^q} \sim \tau^{q/2}$ \citep{Monin1972}, and so the flatness coefficient should remain constant in the inertial range, $\tau_\eta\ll\tau\ll T_L$. Thus, the change of $F(\tau)$ for $\tau\gg\tau_\eta$ shows the existence of deviation from the Kolmogorov similarity theory in the canopy flow. As discussed in Sec.~\ref{sec:intro}, this transition of the statistics with $\tau$ is a hallmark of turbulent flows that characterizes small-scale intermittency.

\cite{Chevillard2003} proposed that the transition from the flat to Gaussian PDF in HIT can be described by the multifractal model, and showed that it was in good agreement with results from two experiments and DNS simulations at various $Re_\lambda$. Briefly explained, in the multifractal formalism the velocity increments are specified as 
\begin{equation}
\Delta_\tau v_i = \mathcal{B}\left(\frac{\tau}{T_L} \right) \Delta_{T_L} v_i
\label{eq:multifractal}
\end{equation} 
where $\Delta_{T_L} v_i$ is the velocity increments at long-times, and $\mathcal{B}\left(\tau/T_L \right)$ is a random function. Then, $P(\Delta_\tau v_i)$ can be calculated by integrating the PDFs of $\mathcal{B}$ and $\Delta_{T_L} v_i$, given a model for $\mathcal{B}$. Importantly, this work uses the same model for $\mathcal{B}$ that was originally utilized by \cite{Chevillard2003} for studies of HIT flows, and it thus assumes the same singularity spectrum for the canopy flow; \rev{a full description of the model is given in the Appendix~\ref{appA}}. The resulting PDFs that were calculated using the model are shown in Fig.~\ref{fig:velocity_increments} as continuous lines underlying the empirical data. The flatness coefficient that was calculated using the multifractal model is also plotted in Fig.~\ref{fig:velocity_increments}(b) as a continuous line, showing a fair agreement between the empirical data and the model. The fair agreement between the empirical results and the model is important because we used here the same function $\mathcal{B}$. Indeed, the fact that using the same singularity spectrum we could obtain a close fit for statistics of our data suggests that there exists a similarity between the small-scale dynamics in the canopy flow and HIT, despite the strong inhomogeneity and anisotropy of the canopy flow.

\begin{figure}
	\centering
	\includegraphics[height=5cm]{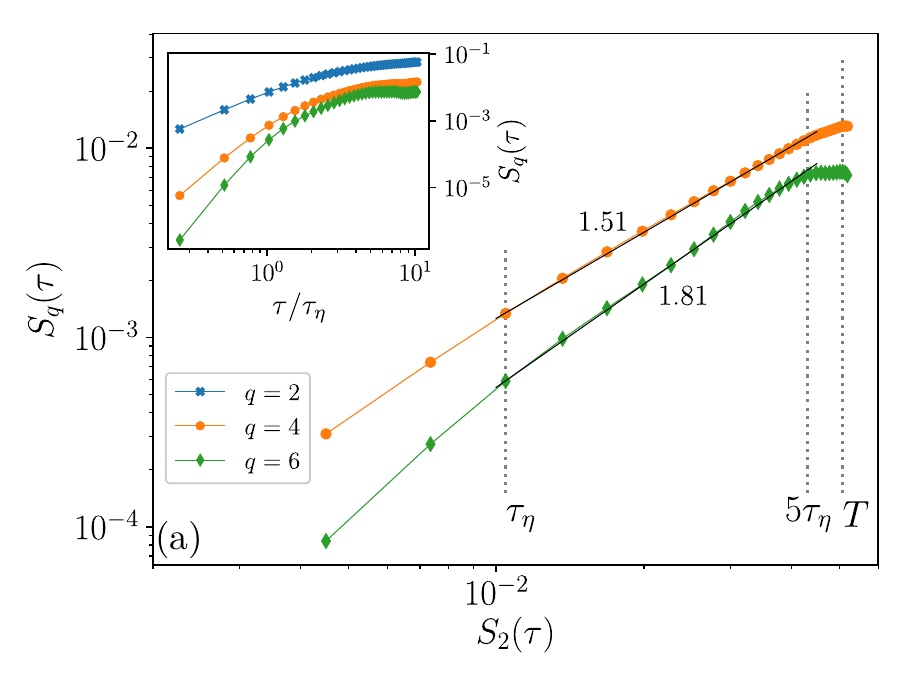}
	\includegraphics[height=5cm]{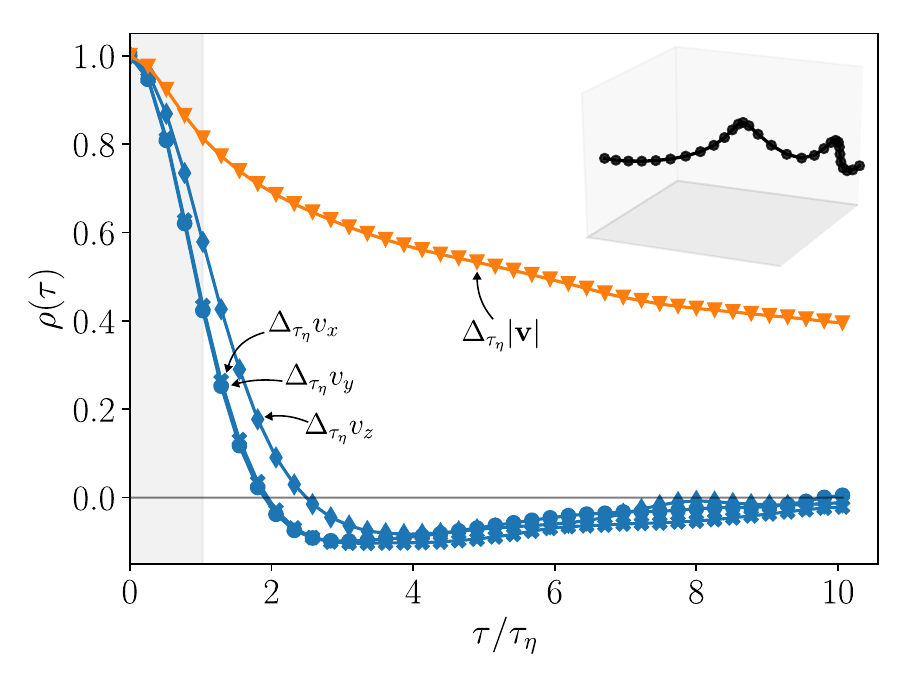}
	\caption{(a) The inset shows Lagrangian structure functions, $S_q(\tau)$, for $q=2$, 4 and 6; the main figure is an ESS plot that shows $S_4$ and $S_6$ against $S_2$ to probe relative scaling. (b) Lagrangian autocorrelation function of temporal velocity increments with $\tau=\tau_\eta$, shown for the \rev{three velocity components} and for the magnitude of the full velocity vector. The inset is a 3D representation of a convoluted trajectory in a box of size $(0.2H)^3$.}
	\label{fig:traj}
\end{figure}

The so-called Lagrangian structure functions are moments of the velocity increments,
\begin{equation}
S_{q}(\tau) = \av{ (\Delta_\tau v_i)^q }\,\, .
\label{eq:Sq}
\end{equation}
\rev{As discussed above, in the so-called inertial range, $\tau_\eta \ll \tau \ll T_L$, the Kolmogorov similarity theory predict that $S_q \propto \tau^{\zeta_q}$ with $\zeta_q = q/2$~\citep{Monin1972}. Here, we would like to examine whether this prediction might hold in the canopy flow as well, while noting that for anisotropic flows like ours we may only speak of "effective" scaling due to effects of anisotropy. Thus, the structure functions for $q=2$, 4, and 6 are shown in log-log scales in the inset of Fig.~\ref{fig:traj}(a). Indeed, no clear scaling region can be found for $\tau > \tau_\eta$ in the graph, however, this is a common feature that occurs in isotropic flows as well. It was suggested that the lack of scaling in isotropic flows may be a result of finite Reynolds number effects~\citep{Toschi2009} and it can hinder theory validations and comparison between different experiments and numerical simulations. A commonly used method to bypass this difficulty is to use the so-called extended self-similarity framework (ESS), in which $\zeta_q$ is examined relative to $\zeta_2$ \citep{Toschi2009}; the ESS approach can extend the scaling range and it was found to successfully converge various previous experimental and numerical results~\citep{Arneodo2008}}.
Thus, in the main panel of Fig.~\ref{fig:traj}(a), we examine $\zeta_q / \zeta_2$ by plotting $S_q$ against $S_2$ in log-log scales for $q=4$ and 6. The figure shows a narrow range $\tau_\eta < \tau \lesssim 4.5\tau_\eta$ in which a scaling exists for the canopy flow experiment. Notably, the separation of scales in the canopy experiment was $\frac{T}{\tau_\eta}\approx6$, which is very low as compared to homogeneous flows with similar $\mathrm{Re}_\lambda$, due to the so-called rapid decorrelation that was explored by \cite{Shnapp2020c}, and this limited severely the extent of the scaling range of $S_q/S_2$. Yet, in the existing range Fig.~\ref{fig:traj}(a) gives $\frac{\zeta_4}{\zeta_2}\approx1.51$ and $\frac{\zeta_6}{\zeta_2} \approx 1.81$. These values are in remarkable agreement with previous experimental results from HIT flows; for example, \cite{Mordant2004} found $\frac{\zeta_4}{\zeta_2}=1.54 \pm 0.06$ and $\frac{\zeta_6}{\zeta_2}=1.8 \pm 0.2$ for the $\mathrm{Re}_\lambda = 570$ experiment (cf. Table 4 there). \rev{The agreement we observe here is important, because it supports the argument of local homogeneity in canopy flows at small scales.}

\rev{Let us briefly consider dynamical scenarios for small-scale Lagrangian intermittency. Results from HIT DNS by \cite{Biferale2005, Bec2006, Bentkamp2019} suggested that small-scale intermittency is a result of encounters between particles and intense vortex filaments; indeed, \cite{Wilczek2008} showed that the characteristic transition of the increments' PDFs can be captured by a heuristic flow model of superimposed constitutive vortices. Similarly, \cite{Liberzon2012} showed that acceleration-vorticity-strain alignment in a quasi-homogeneous flow is associated with intense energy flux. Here, we can show hints suggesting that similar scenarios occur in the canopy flow as well. As shown by \cite{Mordant2002, Mordant2004}, while Lagrangian acceleration components decorrelate on timescales of the order $\sim\tau_\eta$, the magnitude of acceleration decorrelates on much longer timescales. Here, Fig.~\ref{fig:traj}(b) suggests that the same is true in our canopy flow data. It shows four autocorrelation functions: three for the increments of each of the velocity components ($x$, $y$ and $z$) and one for increments of the magnitude of the velocity vector, taking the time lag $\tau=\tau_\eta$ (the velocity difference can be used as a proxy for the acceleration because at such small time-lags the acceleration is still correlated, e.g. see \cite{Voth1998,Shnapp2020c}, so approximately $\Delta_{\tau_\eta}v_i \approx \frac{\partial v_i}{\partial t}\tau_\eta$). While the three components' velocity increments became decorrelated ($\rho=0$) at roughly $\tau\approx2\tau_\eta$, the velocity magnitude difference retained correlation with itself over the whole range of the measurement, with the minimum value of around $\rho\approx0.4$. This difference between the components' and the magnitude's autocorrelations agrees with the vortex trapping picture. In addition to that, for $\tau\gtrsim 2 \tau_\eta$ the components' increments were anti-correlated, which, as shown by ~\cite{Wilczek2008}, can result from trajectories rotation around vortex filaments' cores; this too supports the picture of vortex trapping. Thus, Fig.~\ref{fig:traj}(b) supports the notion that small-scale Lagrangian intermittency in the canopy flow is related to the encounter of trajectories with sparse and intense vortex filaments, similar to the HIT case. The inset of Fig.~\ref{fig:traj}(b) visualizes a convoluted trajectory, which is a possible instance of such a trapping scenario.}

\subsection{Conditional statistics and large-scale intermittency} \label{sec:activity-momentum}

\begin{figure}
	\centering
	\includegraphics[height=5cm]{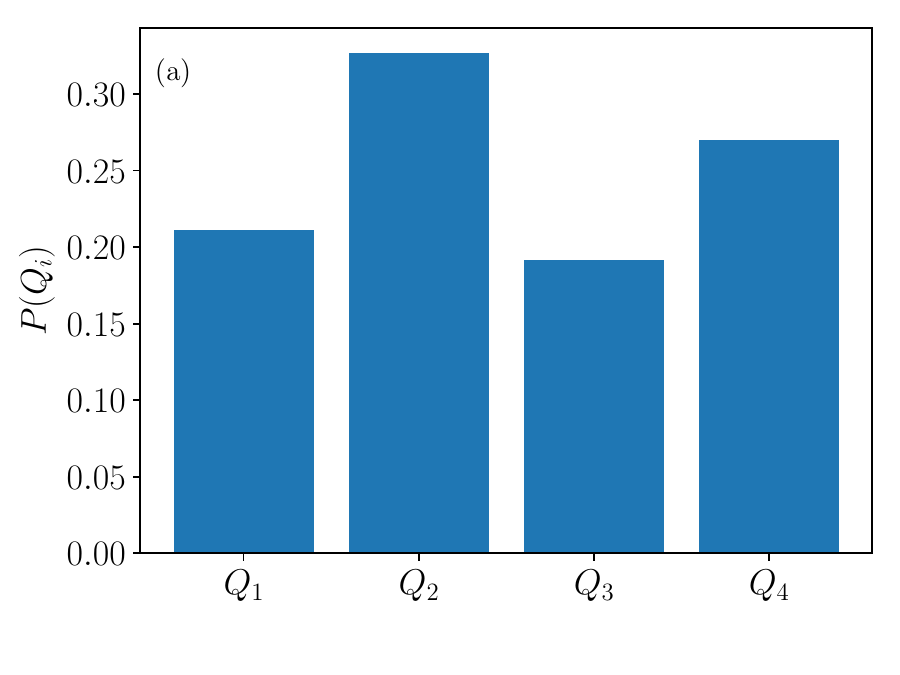}
	\includegraphics[height=5cm]{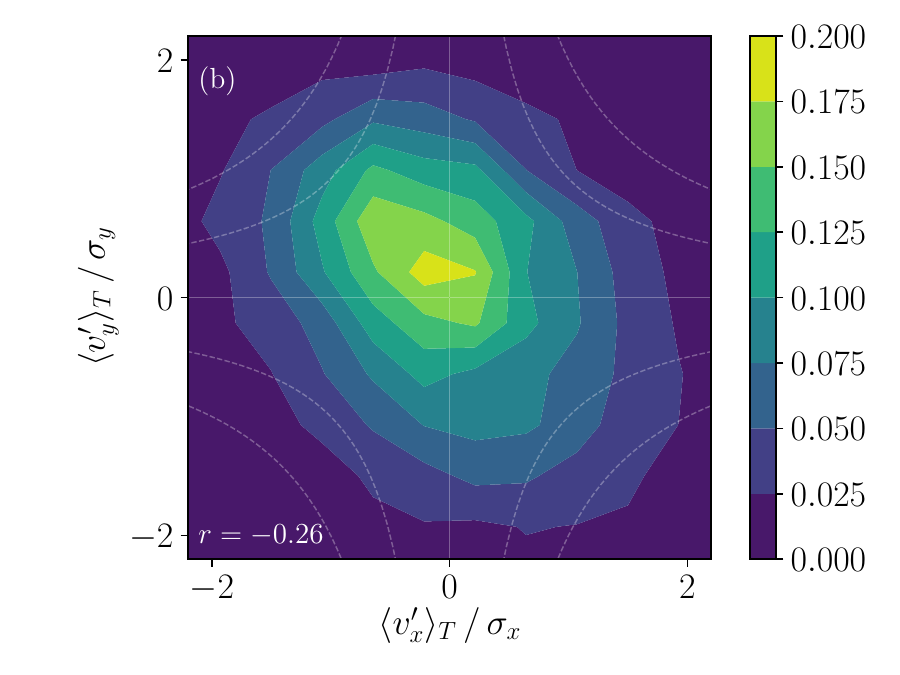}
	\caption{(a) - Normalized histogram for the number of trajectories in our dataset with each quadrant $Q_i$. (b) Joint PDF of the streamwise and vertical velocity components averaged over the velocity decorrelation timescale, $T$. \label{fig:velocity_Q}}
\end{figure}

In the following section we use conditional statistics in order to detect large-scale intermittency. Consider the velocity of a certain Lagrangian trajectory between the times $t_0$ and $t_0+\tau$: $\mathbf{v}_{t_0, \tau} \equiv \{\mathbf{v}(t) \, | \, t_0\leq t< t_0+\tau\}$. The \textit{Lagrangian average} of a function in this section of time shall be denoted with a tilde symbol as
\begin{equation}
\avl{f(\mathbf{v})}_{ t_0, \tau} \equiv \frac{1}{\tau} \int_{t_0}^{t_0 + \tau} f(\mathbf{v}_{t_0, \tau}) dt \,\, .
\label{eq:lag_sec_avg}
\end{equation}
\rev{Note that such averages are properties of individual trajectories over periods of duration $\tau$. Also, since we assume stationarity of the flow we present statistics for different trajectories, namely averaged over $t_0$.} We denote fluctuations of the trajectory averaged velocity with respect to the Eulerian mean velocity as $\avl{\mathbf{v}}_\tau' \equiv \avl{\mathbf{v}}_\tau - \mathbf{U}$. Now, using eq.~\eqref{eq:lag_sec_avg} and in analogy to the Eulerian quadrant analysis \citep{Antonia1981,Shaw1983, Raupach1986, Zhu2007}, we define the \textit{Lagrangian quadrant} of a trajectory using signs of the components of $\avl{\mathbf{v}}_\tau'$ on the $x$ and $z$ plane as follows:
\begin{equation}
Q_i \equiv 
\begin{cases}
1,& \text{if } \quad \avl{v_x}_T' > U_x \,\text{ and }\, \avl{v_z}_T' > U_z\\
2,& \text{if } \quad \avl{v_x}_T' \leq U_x \,\text{ and }\, \avl{v_z}_T' > U_z\\
3,& \text{if } \quad \avl{v_x}_T' \leq U_x \,\text{ and }\, \avl{v_z}_T' \leq U_z\\
4,& \text{if } \quad \avl{v_x}_T' > U_x \,\text{ and }\, \avl{v_z}_T' \leq U_z
\end{cases}
\label{eq:quadrants}
\end{equation}
\rev{where we use the averaging time $\tau=T$, the Lagrangian velocity decorrelation timescale~\citep{Shnapp2020c}.} Fig.~\ref{fig:velocity_Q}(a) shows a normalized histogram for the trajectories being associated with the four quadrant states. It shows that $Q_2$ trajectories were the most common, followed by $Q_4$ trajectories and then $Q_1$ and $Q_3$ trajectories, which is in qualitative agreement with the duration fractions reported by \cite{Yue2007, Zhu2007}. It is also interesting to see that the time averaged Lagrangian velocity fluctuation components are correlated, similar to the Eulerian turbulent velocity components that make up the Reynolds stress. This is shown in Fig.~\ref{fig:velocity_Q}(b) through the elliptical shape of the joint PDF that is elongated in the direction of the negative diagonal; the correlation coefficient was -0.26. Below, conditional statistics based on $Q_i$ are used to probe large-scale intermittency effects on the Lagrangian statistics.

\begin{figure}
	\centering
	\includegraphics[height=5cm]{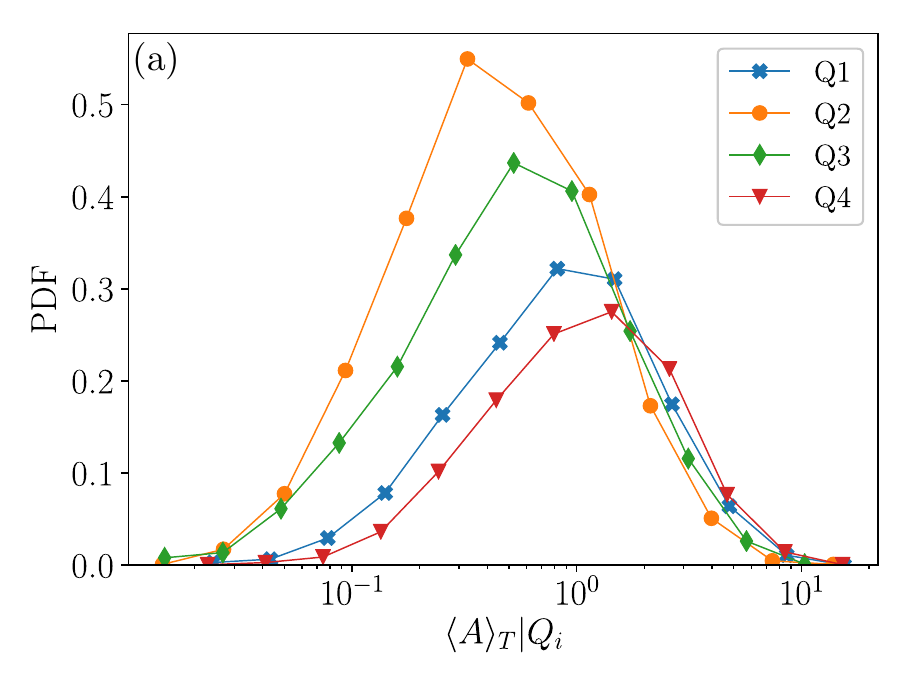}  
	\includegraphics[height=5cm]{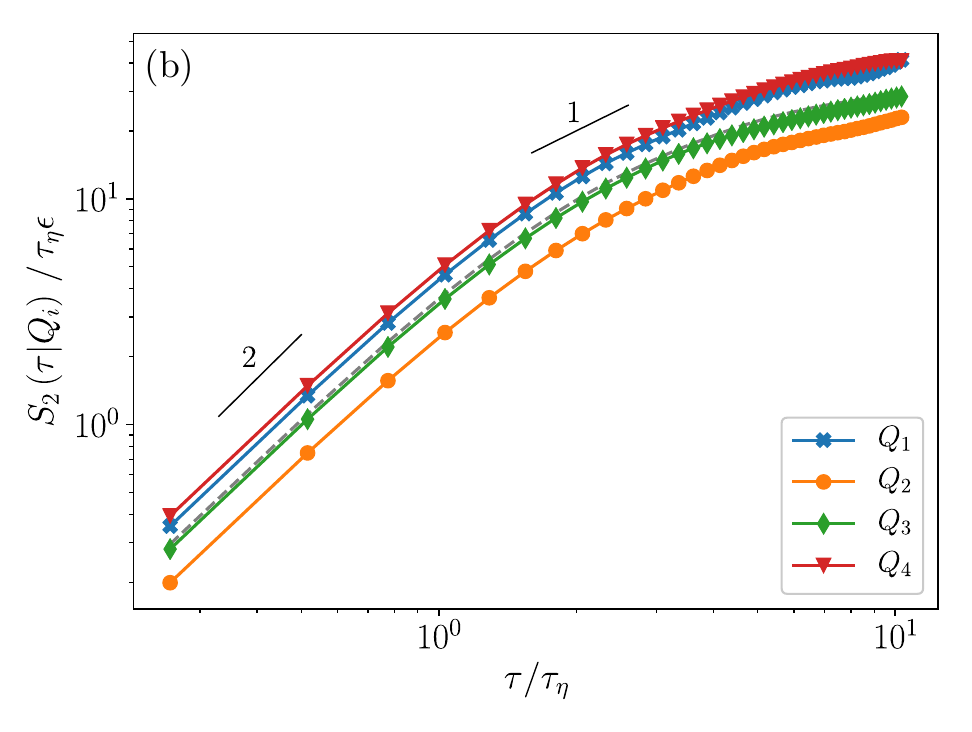}
	\caption{Lagrangian statistics condition with Eq.~\eqref{eq:quadrants}. (a) PDFs of the activity $A_T$ for four groups of trajectories divided according to their quadrant. (b) Second-order Lagrangian structure-function for trajectories from different velocity quadrants. \label{fig:energy_Q}}
\end{figure}

The trajectories in the canopy flow experiment were observed to be associated with more/less strong changes of their kinetic energy ($e\equiv\frac{1}{2}|\mathbf{v}|^2$) when trajectories were conditioned based on the value of $Q_i$. To demonstrate this, let us denote the following property:
\begin{equation}
A_{\tau} \equiv \avl{ \big[e  - E_\tau \big]^2}_{\tau}^{1/2} \,\, ,
\label{eq:activity} 
\end{equation}
where $E_\tau=\avl{e}_\tau$ is the average kinetic energy of a trajectory. Thus, $A_\tau$ is the RMS of the kinetic energy increments that were discussed by~\cite{Xu2014} along the path of a Lagrangian trajectory during a time $\tau$. It is a non-negative scalar that quantifies the amplitude of kinetic energy changes undergone by a trajectory. Loosely speaking, it can be interpreted to show how \textit{active} a trajectory is for fixed durations. \rev{In Fig.~\ref{fig:energy_Q}(a) we show PDFs of $A_{T}$ conditioned on $Q_i$, where note that again we use $\tau=T$}. It is seen that $A_T$ was typically the highest for trajectories with $Q_4$ or $Q_1$, and that it was the lowest for trajectories with $Q_2$. Also, the average of $A_T$ for trajectories with $Q_4$ was more than two times higher than the average over trajectories at $Q_2$, but only 20\% higher than the average over $Q_1$ trajectories. Notably, the PDFs of $A_T$ were roughly log-normal.

Figure~\ref{fig:energy_Q}(a) reveals anisotropy in the kinetic energy increments of Lagrangian trajectories, since statistics of $A_T$ depended on the \textit{direction} of trajectory's velocity fluctuations. While it is expected that statistics of $A_T$ will depend on the magnitude of velocity even in HIT, a dependence on direction reveals a symmetry breaking that can only persist in inhomogeneous or anisotropic flows. \rev{Furthermore, since $A_\tau$ measures Lagrangian fluctuations of the kinetic energy, higher values of $A_\tau$ result from stronger forces that act on particles.} Correspondingly, $A_T$ was typically higher for both $Q_4$ and $Q_1$ which are associated with higher streamwise velocity, whereas the converse occurred for $Q_2$ and $Q_3$ that are associated with lower streamwise velocity (relative to $U_x$). This suggests that the changes in statistics of $A_T$ are due to increased/decreased levels of the canopy drag that fluctuated due to large-scale flow structures in the shear-layer and the boundary-layer above the canopy. This is in qualitative agreement with \cite{Keylock2020} who recently associated streamwise velocity and intermittency in a canopy flow.

\rev{The central role of the energy cascade in the understanding of turbulent flows makes it important to examine how large-scale intermittency is reflected across the different scales. To this end, similar to \cite{Blum2010, Blum2011}, we use conditioned structure functions. Specifically, $S_q$ as defined in Eq.~\eqref{eq:Sq} is now calculated over groups of trajectories with the same $Q_i$.} Fig.~\ref{fig:energy_Q}(b) shows the conditioned $S_2$, plotted on log-log scale. The structure functions for different $Q_i$ have nearly identical shapes, but they are translated vertically with respect to one another. In fact, the structure-functions appear in the figure according to the average levels of $A_T$ observed in Fig.~\ref{fig:energy_Q}(a): $S_2$ is highest for $Q_4$, then $Q_1$, $Q_3$ and the lowest is $Q_2$. In addition to that, since the figure is in log-log scales, the identical shapes mean that the time scaling of structure-functions relating to different quadrants is the same: $\zeta_2\approx2$ for $\tau\leq\tau_\eta$ and it reduces below 1 by the end of our measurement range. In particular, $\zeta_2(\tau)$ was almost independent of $Q_i$.

\rev{The important observation from Fig.~\ref{fig:energy_Q}(b) is that when conditioning on $Q_i$ the changes in the examined statistics occurred homogeneously across the scales. However, results from other flows suggest that this is not a universal feature. For example, \cite{Sreenivasan1998} and \cite{Blum2010, Blum2011} showed that conditioning samples on a representative large-scale velocity affected the scaling of Eulerian structure functions only in some flows while in other flows it did not. While a rigorous explanation of why this occurred in our flow is out of the scope of the present work, we can suggest phenomenological reasoning. In canopy flows, the turbulence is severely obstructed by the canopy obstacles and the forcing acts mostly due to the interaction with fixed boundaries. This is different from free flows, in which turbulence production intrinsically depends on correlations between the flow and the forcing, as shown recently by \cite{Bos2019}. This consideration suggests that energy input occurred mostly on scales larger than our measurement volume, so canopy drag fluctuations did not significantly alter the structure functions on the scales available in Fig.~\ref{fig:energy_Q}(b).} This observation is important for two reasons. First, it implies that changes in statistics when conditioning on $Q_i$ occur due to variations in "turbulence parameters", namely this is indeed large-scale intermittency. Second, it is important for Lagrangian near-field dispersion models since it suggests that temporal fluctuations in canopy drag may be treated by varying the simulation's parameters over long timescales, e.g. as discussed by \cite{Pope1990, Pope1991, Aylor1990, Duman2014, Duman2016}.

\section{Discussion and conclusions}\label{sec:conclusions}

To conclude, this work presents observations of both small-scale intermittency and large-scale intermittency of Lagrangian statistics in a canopy flow by using the results of a recent wind-tunnel experiment. This is the first experimental observation of Lagrangian intermittency in a canopy flow, and thus, it presented a unique opportunity to probe these two different types of intermittency in parallel. Our results demonstrate the importance of direct Lagrangian investigations of inhomogeneous and anisotropic turbulent flows.

The Lagrangian small-scale intermittency was manifested by deviations of the velocity increment's statistics from self-similarity, and in particular, their flatness increased strongly when $\tau$ was decreased. Furthermore, a marked similarity was observed between our results for the canopy flow and previous observations from HIT. Specifically, using Lagrangian the multifractal model and the ESS framework, we found remarkable quantitative agreement with \cite{Chevillard2003} and \cite{Mordant2004}. Lastly, the long correlation of acceleration magnitude and the short correlation of acceleration components suggests that the source for small-scale intermittency is, similar to HIT, rooted in encounters of particles with vortex filaments~\citep{Biferale2005, Bec2006, Wilczek2008, Bentkamp2019}. These results strongly support the picture suggested by \cite{Arneodo2008} of universal Lagrangian intermittency in turbulence, and it also suggests its extension to certain highly turbulent inhomogeneous and anisotropic flows. This observed similarity may have been due to the dominance of the isotropic dissipation terms over contributions from the flow's inhomogeneity to the particle's dynamics, as we reported in \cite{Shnapp2020c}. In this case, the main conclusion is that even in presence of marked inhomogeneity and anisotropy, the HIT picture may still be relevant at small-scales if the turbulence energy flux is sufficiently high.

It was also observed that when conditioned on the direction of the time averaged velocity fluctuation, Lagrangian trajectories had significantly different statistics for the RMS of kinetic energy increments. It was typically much higher (lower) for trajectories whose streamwise velocity component was higher (lower) than the mean. Correspondingly, the second-order Lagrangian structure functions were higher (lower) for these groups of trajectories. This suggests that fluctuations of the canopy drag force affect the activity of Lagrangian trajectories and, therefore, this observation is a manifestation of large-scale intermittency. Furthermore, it was observed that the large-scale intermittency did not affect the scaling of the Lagrangian structure functions, namely that the effect of conditional statistics was felt homogeneously across the different scales. This observation is important for the treatment of large-scale intermittency in Lagrangian dispersion models.\\




\noindent{\bf Acknowledgments\bf{.}} I would like to express my sincere gratitude to Alex Liberzon, Yardena Bohbot-Raviv and Eyal Fattal for the experimental data, and to Narsing K. Jha for enlightening comments.\\





\appendix

\section{The multifractal model}\label{appA}

In the main text, the multifractal model was used to support the argument that small-scale intermittency in the canopy flow reflects processes that are characteristic of fully developed turbulence. The specific formulation of the model we used follows the development by \cite{Chevillard2003}. We only slightly modified it in order to fit the canopy data. According to \cite{Chevillard2003}, the Lagrangian velocity differences are given by - 
\begin{equation}
\Delta_\tau v_i = \mathcal{B}\left(\frac{\tau}{T_L} \right) \Delta_{T_L} v_i
\end{equation} 
where $\mathcal{B}$ is a random function, and their PDF can be calculated as 
\begin{equation}
P(\Delta_\tau v_i) = 
\int_{-1/2}^{+\infty} \frac{\mathcal{P}(h, \frac{\tau}{T_L}, Re, \mathcal{D}(h))}{\mathcal{B}(h, \frac{\tau}{T}, Re)} \mathcal{G}\left( \frac{\Delta_\tau v_i}{\mathcal{B}(h, \frac{\tau}{T_L}, Re)} \right) \, dh
\label{eq:Pdv_t}
\end{equation}
The function $\mathcal{B}$ and its PDF $\mathcal{P}$ were calculated using the exact same specification as in \cite{Chevillard2003}, \rev{and similarly, the PDF of the increments of $\Delta_{T_L} v_i$ was assumed to be Gaussian.} \rev{However, 
	as we discussed in \cite{Shnapp2020c}, while the separation of scales $T/\tau_\eta$ in the the HIT case is a function only of the Reynolds number $T/\tau_\eta = f(\mathrm{Re})$, in canopy flows it depends also on other macroscopic parameters of the flow, such as the geometry or the arrangement of the obstacles. Therefore, to fit the model to the canopy flow data we required an additional parameter, denoted $\vartheta$, that adjusts the separation of scales to the measured values.} Thus, we obtain the following formulation:
\begin{equation}
\mathcal{B}(h, \frac{\tau}{T_L}, Re, \vartheta) = \frac{(\frac{\tau}{T_L} \, \frac{1}{\vartheta} )^h}
{[ 1 + (\frac{\tau}{\tau_\eta(h)})^{-\delta}]^{(1-h)/\delta}}
\end{equation}
and 
\begin{equation}
\mathcal{P}(h, \frac{\tau}{T_L}, Re, \mathcal{D}(h), \vartheta) = \frac{(\frac{\tau}{T_L}\, \frac{1}{\vartheta})^{1-\mathcal{D}(h)}}
{[ 1 + (\frac{\tau}{\tau_\eta(h)})^{-\delta}]^{\left(\mathcal{D}(h)-1\right)/\delta}}
\end{equation}
In addition, the so-called singularity spectrum $\mathcal{D}$, and the local (fluctuating) dissipation timescale were also chosen following \cite{Chevillard2003} as
\begin{equation}
\mathcal{D}(h) = 1 - \frac{(h-c_1)^2}{2(c_1 - \frac{1}{2})}
\end{equation}
\begin{equation}
\tau_\eta = T_L \, Re^{-1/(2h+1)}
\end{equation}

The free parameters of the model are thus the Reynolds number, $\mathrm{Re}$, and the integral timescale, $T$, that were given in section 2 of the paper, and three additional free parameters, $\delta$, $c_1$, and $\vartheta$ that govern the details of the transition from dissipation to the inertial regimes. The three parameters were fitted to the empirical data by minimizing the difference between the flatness coefficient at fixed $\tau$ values. The values that were obtained and that were used to plot Fig.~3 are $\delta=0.6$, $c_1=0.593$ and $\vartheta=3.5$


\bibliographystyle{jfm}
\bibliography{jfm}


	
\end{document}